# The Spectrum of Mechanism-oriented Models for Explanations of Biological Phenomena

## Overview

C. Anthony Hunt[1*], Ahmet Erdemir[2], Feilim Mac Gabhann[3], William W. Lytton[4], Edward A. Sander[5], Mark K. Transtrum[6], and Lealem Mulugeta[7]

[1] Department of Bioengineering and Therapeutic Sciences, University of California, San Francisco, CA
[2] Department of Biomedical Engineering, Lerner Research Institute, and Computational Biomodeling Core, Lerner Research Institute, Cleveland Clinic, Cleveland, OH
[3] Institute for Computational Medicine and Department of Biomedical Engineering, Johns Hopkins University, Baltimore, MD
[4] Departments of Neurology and Physiology, and Pharmacology, SUNY Downstate Medical Center; Department Neurology, Kings County Hospital Center, Brooklyn, NY
[5] Department of Biomedical Engineering, University of Iowa, Iowa City, IA
[6] Department of Physics and Astronomy, Brigham Young University, Provo, UT
[7] *InSilico* Labs LLC , Houston, TX

* Corresponding author: a.hunt@ucsf.edu

## ABSTRACT

Within the diverse interdisciplinary life sciences domains, semantic, workflow, and methodological ambiguities can prevent the appreciation of explanations of phenomena, handicap the use of computational models, and hamper communication among scientists, engineers, and the public. Members of the life sciences community commonly, and too often loosely, draw on "mechanistic model" and similar phrases when referring to the processes of discovering and establishing causal explanations of biological phenomena. Ambiguities in modeling and simulation terminology and methods diminish clarity, credibility, and the perceived significance of research findings. To encourage improved semantic and methodological clarity, we describe the spectrum of Mechanism-oriented Models being used to develop explanations of biological phenomena. We cluster them into three broad groups. We then expand the three groups into a total of seven workflow-related model types having clearly distinguishable features. We name each type and illustrate with diverse examples drawn from the literature. These model types are intended to contribute to the foundation of an ontology of mechanism-based simulation research in the life sciences. We show that it is within the model-development workflows that the different model types manifest and exert their scientific usefulness by enhancing and extending different forms and degrees of explanation. The process starts with knowledge about the phenomenon and continues with explanatory and mathematical descriptions. Those descriptions are transformed into software and used to perform experimental explorations by running and examining simulation output. The credibility of inferences is thus linked to having easy access to the scientific and technical provenance from each workflow stage.

# INTRODUCTION

The use of "mechanistic model" and similar phrases in life sciences research literature continues to increase (Figure 1).  However, there is considerable diversity in what is being implied when discussing mechanisms and/or describing models as mechanistic.  Mechanistic model is a convenient yet ambiguous phrase typically used as an abbreviation for more accurate, more informative descriptors.  Use of the term "mechanism" is often similarly ambiguous.  In any scientific community, clarity within research reports and credibility of claims made are generally viewed as being correlated, and computational biology is not an exception.  Usage of ambiguous phrases within simulation research reports enhances the impression of inaccessibility, which can limit the credibility and acceptance of evidence and insights being presented within those reports.  This overview is motivated by ongoing collaborative efforts to improve credibility and the belief that improvements in semantic and methodological clarity will strengthen the credibility of results leveraging simulation research.

The phrase "mechanistic model" has a variety of meanings ascribed to it that differ across biological domains.  There is an increasing tendency to utilize "mechanistic model" both specifically and as an umbrella term.  Herein, we define, characterize, and cluster seven mechanistic model types, and suggest specific terms for each.  To insure clarity, we narrow the scope of discussions that follow by first limiting attention to reports seeking mechanism-oriented explanations of biological phenomena.  We further restrict focus to research for which a scientific objective is to 1) provide deeper, more explanatory insight into the generation of biological phenomena; and/or 2) better predict, mimic, or emulate one or more biological phenomena.

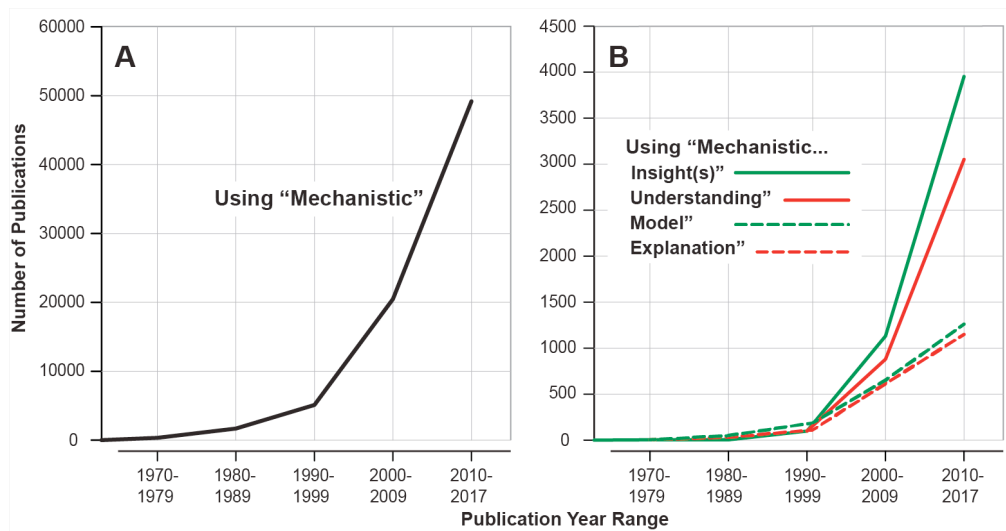

**Figure 1: Use of the term "Mechanistic" in literature**
There is a rapid increase in the use of (A) the term "mechanistic"; and (B) derivative phrases (five shown) in biomedical literature.  Results are from a PubMed search conducted on August 29, 2017.

The purpose of this overview is to illustrate specific ways in which semantic and methodological clarity regarding mechanisms and explanations of phenomena can be refined to improve accessibility, and ultimately, credibility.  A goal is to clarify various uses of "mechanistic model" and how they are represented computationally for explaining biological phenomena.  We describe the spectrum of Mechanism-oriented Models being used to develop explanations of biological phenomena. We cluster explanations of phenomena into three broad groups and then expand them into a total of seven model types. We name each type and illustrate with diverse examples drawn from the literature. We begin by framing the context and offering definitions. In "Methodological Complexity," we contrast how infrastructure and management of complexity influence clarity differently between wet-lab and simulation research. In the section that follows, we describe three spectra that are useful in describing, characterizing, and distinguishing explanations of phenomena. Next, in "Three Groups of Models of Explanation," we use similarities and differences (with reference to the spectra) to guide characterizations that distinguish semantically among seven workflow-centered models of explanation, including four different types of computational models of explanation. The names used to identify each characterization are not intended as semantic standards; rather they are offered as suggestions to encourage movement in that direction and serve as a working foundation for an ontology to use in explanatory simulation research in the life sciences. In "Relevant Information, Multiple Sources," we illustrate why providing sufficient information is essential to enhance credibility of an explanatory simulation, whereas brevity weakens credibility at the expense of clarity. We characterize five different sources and types of information from which relevant details are needed to clearly distinguish among the four types of computational models of explanation. In "Workflow, Provenance, and Hybrid Models," we comment on connections between workflows and semantics, and on new technical issues that further increase the need for semantic clarity, which is followed by "Concluding Remarks."

## BACKGROUND

**Framing the Context: Mechanisms as Explanations of Phenomena**

A prerequisite for discussing mechanism-oriented biological models is adopting a definition for "mechanism." Over the past two decades, within the philosophy of science literature, mechanism has emerged as a framework for thinking about fundamental issues in biology [1, 2].

Braillard and Malaterre recently defined a biological mechanism [3]:

> *"A mechanism can be thought of as being composed of parts that interact causally (usually through chemical and mechanical interactions) and that are organized in a specific way. This organization determines largely the behavior of the mechanism and hence the phenomena that it produces. ... Mechanisms can be formalized in different ways, including with the help of diagrams and schemas, and are usually supplemented by causal narratives that describe how the mechanisms produce the very phenomena to be accounted for."*

Authors often augment their diagrams, schemas, and causal narratives with a computational "narrative" (algorithm and implementation) that enables explicit predictions. We use the definitions listed in Working Definitions Sidbar and specify that a mechanism is a real thing; it is concrete. A description is required for the term "mechanistic model." Kaplan and Craver state [4]:

> *"[That] the line that demarcates [mechanistic] explanations from merely empirically adequate models seems to correspond to whether the model describes the relevant causal structures that produce, underlie, or maintain the explanandum phenomenon. This demarcation line is especially significant as it also corresponds to whether the model in question reveals (however dimly) knobs* *and levers that potentially afford control over whether and precisely how the phenomenon manifests."*

Thus, we see that there is a difference between a model that reproduces a phenomenon and a model that does so using a mechanism that recapitulates the 'true' underlying mechanism.

---

**(Sidebar) Working Definitions:**

**mechanism** n : 1) a structure, system (e.g., biological, mechanical, chemical, electrical, and so on), or process performing a function in virtue of its component parts, component operations, and their organization (adapted from [5]), where the function is responsible for the phenomenon to be explained; 2) entities and activities organized in such a way that they are responsible for the phenomenon to be explained (adapted from [6,7])

**phenomenon** n : 1) an observable fact or event: an item of experience or reality; 2) a fact or event of scientific interest susceptible of scientific description and explanation [8]

**mechanistic** adj : 1) determined by, for example, a mechanical, chemical, and/or electrical mechanism, or executing software; 2) like, for example, a mechanical, chemical, or electrical mechanism in one or more ways; 3) of or relating to using a mechanism as an approach to explaining a biological phenomenon; 4) mechanism-oriented

---

Craver posits that mechanistic models are explanatory, but he notes [9]:

*"Some models sketch explanations but leave crucial details unspecified or hidden behind filler terms. Some models are used to conjecture a how-possibly explanation without regard to whether it is a how-actually explanation."*

The increasing variety and sophistication of published mechanism-oriented and mechanism-based explanatory models reflect that biological mechanisms exhibit features that are not expressed in the Working Definitions Sidbar definition of mechanism. Darden discusses how features of mechanisms often become necessary parts of adequate descriptions and representations of a mechanism [10]. She identifies five features of biological mechanism, listed in Table 1, that often characterize mechanisms that adequately explain biological phenomena. These features will be useful in broadly distinguishing among model types and may provide a basis for further developing an ontology to support mechanism-oriented simulation research. The phenomenon to be explained is the first feature because the search for a mechanism-based model of explanation requires that the phenomenon be clearly identified. Also, in biology, it is often the case that phenomena at a finer biological scale constitute the explanatory mechanism of the phenomenon of interest observed at coarser biological scale. Stated differently, the underlying finer details are the entities and activities responsible for observable coarser behavior.

**Table 1** – Five features of a biological mechanism (adapted from [10]): a biological mechanism exhibits all five.  A computational mechanism-based model may strive to do the same.

**Mechanism**

| Features | Examples | Explanations |
|---|---|---|
| Phenomenon | | A clearly identified phenomenon is the requisite for specifying the other four features of mechanism and for developing a credible explanation of that phenomenon. |
| Components | entities, activities, modules, processes, underlying finer details | Working entities act in the mechanism.  Activities are producers of change.  Some entities and activities can be organized into a module.  Inner layer phenomena can be the entities and activities responsible for the outer layer phenomenon. |
| Spatial arrangement of components | localization, structure orientation, connectivity, compartmentalization | Components are typically localized and organized into a structure.  A component's orientation can be a prerequisite for an activity.  Producing change requires connectivity.  Compartmentalization facilitates spatial arrangement within a structure. |
| Temporal aspects of components | order, rate, duration, frequency | Entities may play their role is a particular order.  Some activities have characteristic rates.  Activities can occur in stages and/or exhibit temporal organization.  An activity and/or stage can repeat or exhibit frequencies.  Stages can unfold in a particular order and have duration. |
| Contextual locations | location within a hierarchy and/or within a series | A mechanism is situated in wider context, such as within a hierarchy of mechanism levels or within a temporal series of mechanisms not directly influencing the phenomenon of interest. |

## METHODOLOGICAL COMPLEXITY

Methodological complexity has been increasing in wet-lab research for decades.  However, for wet-lab researchers, striving for clarity in descriptions of experiments is an ingrained best practice, although one that is not entirely fulfilled in practice.  Ironically, though it is possible to fully document every aspect of software used, such clarity is not yet the norm in the computational biology research domain.  Clarity in reports of wet-lab methods is facilitated and enabled by a large, trusted commercial infrastructure.  Research reports can achieve clarity without having to include pages of essential yet mind-numbing details by including statements like the following within Methods sections, e.g., from [11]:

> *"Dulbecco's phosphate buffered saline (PBS), liver perfusion medium, hepatocyte wash medium … were purchased from Life Technologies (Carlsbad, CA) … Wild-type C57BL/6J, male mice (9 weeks of age), purchased from The Jackson Laboratory (Bar Harbor, ME), were acclimated … The resulting supernatant was injected into the high-performance liquid chromatography column using a Model 582 solvent delivery system and a Model 5600A CoulArray detector (ESA,*

*Chelmsford, MA) … Protein content was determined using the Nanodrop 2000 Spectrophotometer (Thermo Scientific, Waltham, MA)."*

For each item, additional details are available on the manufacturer or supplier's websites. Also, many portions of wet-lab protocols are replicated from previous publications in which each step was explicated, e.g., "cell toxicity was measured as in [hypothetical reference]." There are even entire journals devoted to the distribution of standardized and generalizable protocols, e.g., "Journal of Visualized Experiments" and "Nature Protocols." By contrast, in biology simulation research, particular computational methods are often borrowed and repurposed but are rarely implemented and/or executed identically. Proprietary and open source simulation tools and packages are available, but we do not yet have commercial infrastructure specifically intended to facilitate biology simulation research.

Growth and diversification of the commercial infrastructure supporting biology research have been fueled in part by the requirement that, when needed, experiments can be independently reproduced and extended in a different laboratory. That requirement also drives the need for clarity in wet-lab methods. Interest in independently reproducing results of simulation experiments, and in reusing and repurposing simulation components is expected to increase as the healthcare implications and benefits of simulation experiments increase. Improved clarity at all workflow stages will facilitate those developments.

## MECHANISM-ORIENTED MODELS OF EXPLANATION

Based on our sampling of the research literature, all explanations of phenomena that draw on features of mechanisms can be broadly described as being mechanism-oriented models of explanation. They differ from other models of explanation in that they try to organize knowledge about both phenomenon and its explanation around mechanisms [3]. The explanations are models because, even when there is considerable knowledge about a phenomenon, there is still uncertainty about details of the actual causal process, and those details always exhibit biological variability. They range from being mechanism-oriented to fully mechanism-based models of explanation, as illustrated by the spectrum in Figure 2A, and can be grouped under one of three broad characterizations. **I**: The details of the explanation are mechanism-oriented, but fall short of the definition in the Working Definitions Sidbar. **II**: The explanation is mechanism-based in that it builds on a description of a mechanism that meets the definition of mechanism in Working Definitions Sidbar. However, the mechanism is an analogy based most often on a real or hypothetical engineering, physical, mechanical, chemical, and/or electronic mechanism. **III**: The details of the mechanism-based explanation strive to be biomimetic, not analogical: some entities and activities map directly to biological counterparts. In the next section, we explain and elaborate these three characterizations, extend them to include four types of computational models of explanation (**IV-VII**), and present examples (the seven Roman numerals refer to the names of model types characterized below in Group A, B, and C subsections).

**II** and **III** have two requisites: there must be a clear mapping between the representation of entities and activities, and the target (referent) phenomenon, and the phenomenon must be specified clearly. Phenomena are grounded to the particular experiments or clinical trials in which they were observed and measured. In research, knowledge of a phenomenon can vary dramatically, yet there is a direct relationship between what is known about the phenomenon and the extent to which a mechanism-oriented model of explanation can become sufficiently accurate.  The scope and depth of knowledge about a phenomenon can be characterized by an approximate location along the spectrum in Figure 2B.  Phenomena that are the focus of more basic research tend to have central or left-of-center locations.  A mechanism-oriented or mechanism-based explanation of how a phenomenon is thought to be—or might be—generated can be characterized by an approximate location along the spectrum in Figure 2C.  Photosynthesis provides an example where the explanation of the phenomenon is located right of center on the Figure 2C spectrum.  The depth of knowledge is such that explanatory mechanisms described in review articles and textbooks are broadly accepted as accurate, even though they fall far short of a full and complete account of what actually occurs in a particular plant under particular conditions.  The more detailed descriptions include all features listed in Table 1.  As such, it is accurate to describe such an explanation as a Model Mechanism.

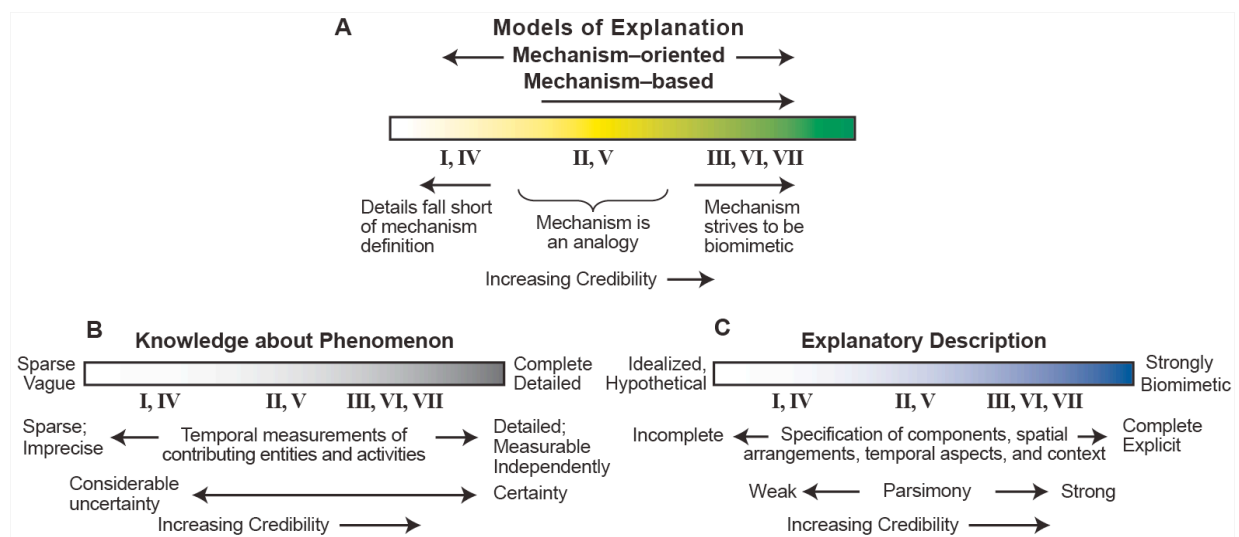

**Figure 2: Three spectra for characterizing the explanation of a phenomenon**
(A) This spectrum illustrates relative relationships among the three types of Mechanism-oriented Models of explanation (**I–III**) illustrated and named in Figure 3, along with two groups of computational models of explanation illustrated and named in Figures 4 (**IV–V**) and 5 (**VI–VII**).  There is often a correlation between characterization and locations on this spectrum and location on spectra B and C.  For example, having locations on B and C that are right of center enables an Analogous-mechanism Model to be more biomimetic.  Explanations that use mechanism analogies often have more centric locations on B and C.  (B) Specifying an approximate location on this spectrum provides a clear, relativistic assessment about the strength of knowledge and information that is available to characterize the phenomenon.  Independent of location, credibility is increased by making explicit information on 1) how the phenomenon has been measured, along with 2) details about temporal measurements of entities and activities thought to be contributing to its generation.  Assessments of uncertainties further increase credibility.  (C) Specifying an approximate location on this spectrum characterizes what is currently known or hypothesized about how the phenomenon in B may be (or is) generated along with information about the mechanism features listed in Table 1 and their orchestration.  Making relative spectra-location information explicit is essential for increasing credibility.

Autoprotection is described as resistance to toxicant re-exposure following acute, mild injury with the same toxicant, such as acetaminophen [12, 13]. It is an example of a phenomenon that can be characterized as located on the far left of the spectrum 2B.  It Knowledge of the phenomenon is sparse and imprecise.  Although there is considerable information about particular molecular details, only incomplete speculative explanations of the phenomenon are currently feasible, and it would be difficult to distinguish causes from effects.  Such explanations would fall short of the definition of mechanism, and so would be located considerably left of center on the spectrum 2C.  As such, weak Mechanistic Explanation is an accurate descriptor, and any possible mechanism-based account would be at best conjecture.

## THREE GROUPS OF MODELS OF EXPLANATION

A huge variety of explanatory model types populates the Mechanism-oriented Models spectrum in Figure 2A.  Having characterizations and descriptors that make it easier to distinguish among classes and types is essential to support clarity and credibility, aid in distinguishing among computational model types, and provide a foundation for an ontology.  We identify and describe seven broad types and cluster them into three groups.  Group A includes the three characterizations illustrated in Figure 3.  One of those characterizations is a requisite core component of each of the four computational Mechanism-oriented Models illustrated in Figure 4 (elaborations of **I** and **II**) and Figure 5 (elaborations of **III**).  As the descriptors and names for different models of explanation gain traction, attention can turn to discussions of finer grain model types, possibly drawing on features listed in Table 1.

**Group A: three types of Mechanism-oriented Models of Explanation**

*I: Mechanistic Explanation*

Mechanistic explanations are pervasive in the life sciences research literature.  In their simplicity, they are analogous to a cartoon; they are static and reflect observations.  Knowledge about the phenomenon is characterized by a location considerably left of center in spectrum 2B and is insufficient to meet the definition of mechanism in the Working Definitions Sidbar.  Nevertheless, there is often sufficient information to support an incipient coarse grain causal story that accounts reasonably well for the available evidence and explains how the phenomenon might have been generated.  The blue box in Figure 3A represents the workflow required to identify and organize relevant information into a description of how the phenomenon might be generated.  Such descriptions typically rely heavily on explanatory diagrams.  They may also include mathematical descriptions, but they fall short of the definition of mechanism, which is clear in the three examples that follow.  It is understood, but often not stated, that many somewhat different, yet equally possible explanatory models can be presented.  An accurate descriptor is a Mechanism-oriented Model of Explanation.  However, because we use that phrase as an umbrella expression, we prefer the abridged phrase, Mechanistic Explanation, which we use hereafter.

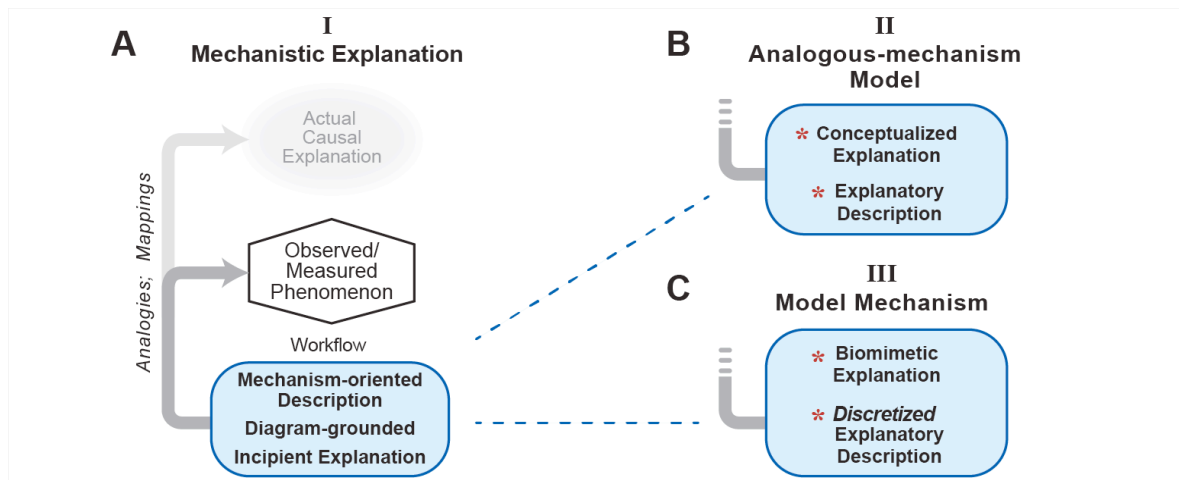

**Figure 3: Three types of Mechanism-oriented Models of Explanation**
There are three broad types of Mechanism-oriented Models of explanations of phenomena. They overlap to some extent. These illustrations highlight features that differentiate the three types. Credibility improves by making clear which type best characterizes a specific model of explanation. (A) A Mechanistic Explanation has the illustrated features and is located left of center in Figure 2A. The muted oval at top, which is repeated in Figures 4 and 5, reminds us that the actual causal explanation is yet to be discovered. The hexagon depicts the target phenomenon. Each phenomenon can be characterized by its relative (to other phenomena) location on the Figure 2B spectrum. The hexagon represents the organized relevant information about the phenomenon that is being explained. The process (the workflow) of identifying and organizing information and features into a description of how the phenomenon might be generated is represented by the blue box. Part of the workflow involves establishing mappings and drawing analogies between features of the explanation and particular measurements; the darker gray arrow indicates that activity. The lighter gray arrow indicates a working hypothesis, in which those mappings and analogies will eventually be extend to the actual causal explanation. Those three differentia are part of the model. (B) This model type includes a detailed description and explanation (blue box) along with the other elements in A, as implied by the incomplete gray arrow. Information about possible generators (of the target phenomenon) is sufficient to conceptualize and describe an explanation that meets the definition of mechanism in the Working Definitions Sidbar by drawing on analogies to, for example, engineering, mechanical, chemical, or electronic mechanisms. The result is an Analogous-Mechanism Model of explanation. The red asterisks designate characteristics that distinguish **II** from **I**. (C) Further right on both the 2B and 2C spectra, knowledge about the phenomenon is sufficient to conceptualize a model of explanation that includes several of the Table 1 explanatory biomimetic features. The resulting detailed description and explanation is fundamentally different from **II**: it is a description of a Model Mechanism explanation. It includes a more detailed description and explanation (blue box) along with the other elements in A, which is implied by the incomplete gray arrow. The red asterisks designate characteristics that distinguish **III** from **II**.

- *Example I-1*: The well-known Hodgkin and Huxley model is a Mechanistic Explanation. It is an incomplete how-possibly story that provides preliminary insights into mechanisms responsible for generating and propagating action potentials along axons [14]. The authors make clear that their account is merely an explanatory model, not an actual explanation.

> *"…certain features of our equations were capable of a physical interpretation, but the success of the equations is no evidence in favour of the mechanism of permeability change that we tentatively had in mind when formulating them."*

- *Example I-2*: Russmann et al. [15] offer a three-step Mechanistic Explanation of how hepatocyte death may be caused during drug induced liver injury. 1) The initial injury results in direct cell stress possibly including mitochondrial impairment. 2) Death receptor-mediated pathways are triggered leading to mitochondrial permeability transition. 3) The result is apoptotic or necrotic cell death.

- *Example I-3*: Bassler et al. [16] sought Mechanistic Explanations for unanticipated clinical side effects and efficacy limitations of integrin αIIbβ3 antagonists. They posited a three-stage Mechanistic Explanation involving paradoxical platelet activation by αIIbβ3 antagonists: a ligand-bound conformation change; receptor clustering; and pre-stimulation of platelets.

*II: Analogous-Mechanism Model*

It is common to encounter a mechanism-oriented explanation of biological phenomenon that is framed as a mechanism analogy based on engineering principles, continuum mechanics, chemistry, electronics, etc. When the analogical explanation meets the definition of mechanism, it can be accurately identified as an Analogous-Mechanism Model of Explanation (simply Analogous-Mechanism Model hereafter). It too is supported by diagrammatic depictions and often includes mathematical descriptions. Like **I**, it is still cartoonish. There is more cause and effect than in **I**. Although the mechanism's phenomenon is expected to be biomimetic, features of the mechanism's components, their spatial arrangement, and/or temporal aspects are typically not biomimetic. The following are examples.

- *Example II-1*: The three-element Hill muscle model for estimation of muscle force generation [17] is an idealized Analogous-Mechanism Model (Figure 3B). Such models do not have direct biological counterparts, and any contextual location is hypothetical. However, measurements of the idealized mechanism during operation—if it were made real, concrete—are expected to adequately match measurements of the target phenomenon qualitatively and quantitatively.

- *Example II-2*: Some therapeutic proteins such as trastuzumab, which is a monoclonal antibody, bind to pharmacological targets on cells. Efficacy is disrupted when the therapeutic protein binds instead to soluble targets shed from cells. Li et al. [18] describe a minimal physiologically based pharmacokinetic Analogous-mechanism Model intended to represent key features of a plausible mechanism hypothesized to be responsible for reduced efficacy. A computational description of their model in operation was used to simulate efficacy changes.

- *Example II-3*: Demographic collapse of freshwater fish species, such as brown trout, can occur when rates of environmental change exceed the population's capacity to adapt. Ayllón et al. [19] describe a spatially explicit, multi-attribute, eco-genetic individual-based Analogous-mechanism Model that was used to study possible trout dynamics under three scenarios: 1) climate change-induced warming, 2) warming plus flow reduction resulting from climate and land use change, and 3) a baseline of no environmental change.

A phenomenon that is explained using an Analogous-mechanism Model will be to the right of **I** in Figure 2B. As explanatory insight improves and the research workflow advances, one encounters research reports in which an earlier Mechanistic Explanation is replaced by an Analogous-mechanism Model. At that stage, authors typically assign names to some or all of the components of their model that are identical to real components and features of the referent biological system, i.e., they draw directly from vocabularies of anatomical or physiological ontologies. While conceptually useful, such labeling can encourage conflating model explanation features with reality, which reduces both clarity and scientific credibility.

*III: Model Mechanism*

As explanatory knowledge about a phenomenon increases (moving further right on the Figure 2B spectrum), researchers begin conceptualizing and describing (hypothesizing about) a particular mechanism-based explanation of the phenomenon (Figure 3C) that is biomimetic; it is not an analogy of something else. Researchers strive to identify, specify, and characterize some or all of the explanatory features in Table 1. Model Mechanism is an accurate descriptor of the product of that process. Model Mechanisms are less cartoonish than **II** and more structured. An early stage model of explanation of this type would likely be assigned a central location on spectrum 2A. As a description matures, its location on all three Figure 2 spectra shift rightward. Mappings exist between the Model Mechanism's discrete entities and activities, and biological counterparts. The expectation is that measurements of a phenomenon generated during simulation of an actualized Model Mechanism would adequately match measurements of the actual target phenomenon qualitatively and quantitatively.

- *Example III-1*: An illustrative example is the two-dimensional model mechanism developed by Norton et al. [20] to facilitate achieving two related goals: 1) improve explanatory insight into the generation of the four distinguishable morphologies of ductal carcinoma in situ of the breast. 2) Disentangle the mechanisms involved in tumor progression. Additional examples are included with those provided below under Group C.

**Group B: using simulation to support and enhance** I **and** II

*IV: Simulation of a Mechanistic Explanation*

A frequent simulation research goal is to translate a Mechanistic Explanation (**I**) into simulation output that is (or is expected to be) qualitatively or quantitatively similar to reported measurements of the target phenomenon. An additional goal may be providing predictions and/or further improving insight into how the target phenomenon (and possibly other phenomena) may be generated. A Simulation of a Mechanistic Explanation (Figure 4A) builds upon **I** during three workflow activities. 1) Relational and continuum mathematical descriptions are developed of the salient explanatory information within the Mechanistic Explanation. 2) Those descriptions are instantiated in software; features to facilitate exploratory simulations are added; solvers are selected, and the implementation undergoes verification. 3) An iterative workflow process achieves the desired qualitative and quantitative similarity between simulation output and measurements of the target phenomenon. During that process, the implemented model and mathematical descriptions may be revised. To enable another modeler to independently reproduce reported simulation results, details of those workflow decisions should be made available when results are published [21].  For the second and third activities, it is increasingly common for researchers to rely on mathematical modeling tools, such as Matlab (The MathWorks, Inc., Natick, MA), and/or proprietary or open source systems, including, for example, physiologically based simulation or emulation packages (e.g., see [22]).  Use of standardized software increases credibility, reliability, and reproducibility by providing some assurance that the underlying numerical techniques are handled correctly.  Use of open source software further improves reproducibility by making the simulation widely available while also opening the underlying techniques to later examination for correctness.

Technically, the simulation output is a model of solutions to the relational and mathematical descriptions under particular conditions; and the mathematical descriptions are a model of the mechanistic explanation in **I** given particular assumptions and constraints.  Consequently, when "mechanistic model" or "mechanistic simulation model" is used to describe the work product, it can be difficult for a reader to know which model is being identified.  To avoid misinterpretations, this type of work product can be identified accurately as a Simulation of a Mechanistic Explanation.  The following are two related examples.

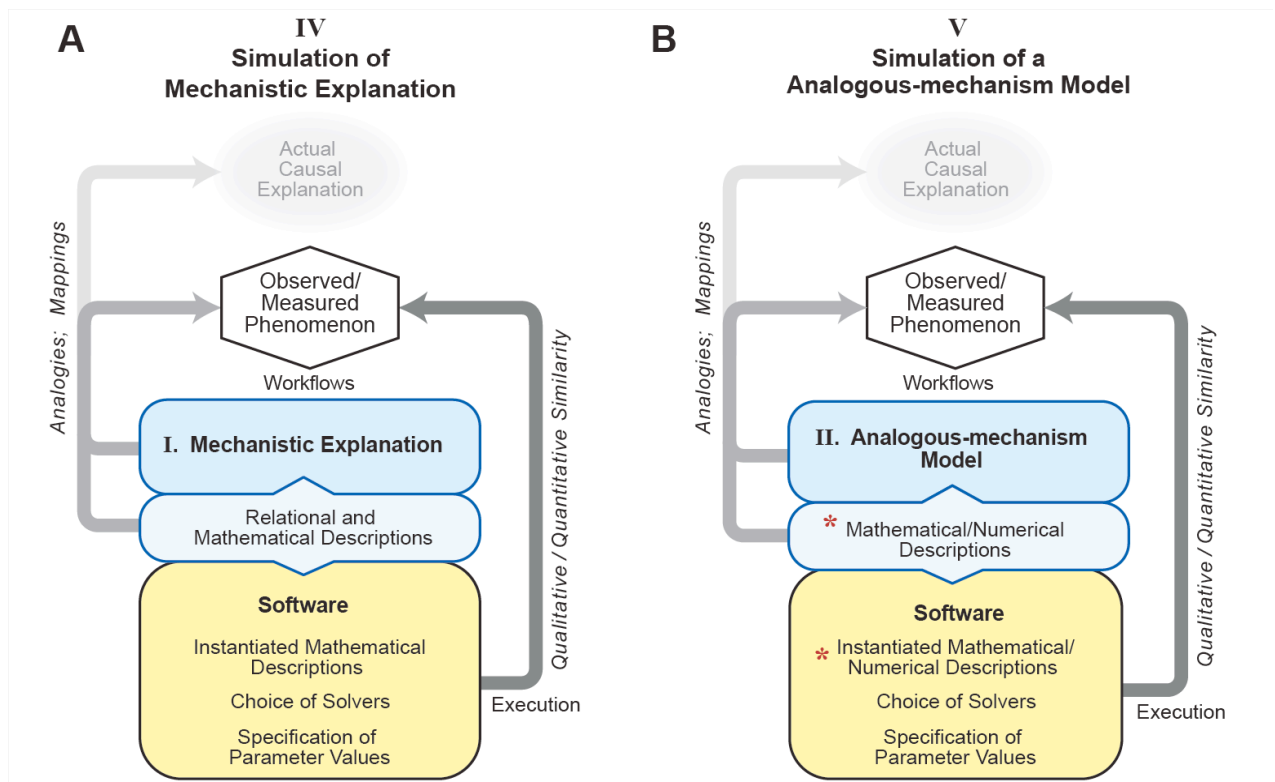

**Figure 4: Characterizations of two types of simulation**
Illustrated are work activities built upon explanations carried forward from **I** and **II**.  Simulation operation is not illustrated.  A requirement for both types of simulation is that output (specific computed solutions) match target phenomenon measurements within some tolerance.  (A) Starting with a Mechanistic Explanation (**I**), the modeler completes two tasks. 1) Develop relational and continuum mathematical descriptions of the mechanistic explanation's salient information.  2) Faithfully instantiate in software all mathematical descriptions such that computed solutions simulate the output envisioned by a) those mathematical descriptions and b) the mechanistic explanation's salient information. The resulting software system provides a Simulation of a Mechanistic Explanation. Prior to publication, the system has typically undergone several rounds of refinement and revision. (B) Starting with **II**, the modeler develops the mathematical and numerical mathematical descriptions needed to provide faithful characterizations of the analogous mechanism's salient features during operation. The requirements for software instantiation are the same as for A. The resulting system simulates output from **II**, as if it were real. Red asterisks: characteristics that distinguishes B from A.

- *Example IV-1*: The gamma rhythm is one of several characterized oscillations of activity in the brain (brain waves). The alpha rhythm of about 8 Hz is powerful enough that it can be readily detected outside of the head, something discovered in the 1920s by Hans Berger. In contrast to alpha, gamma oscillations are faster (~40 Hz) and more spatially localized, best detected by electrodes placed directly on the brain surface or into the brain parenchyma. A Simulation of a Mechanistic Explanation [23] helped explore how these gamma oscillations could be generated through inhibitory inputs, which were classically thought of as delaying or eliminating neural activity. Wang and Buzsaki demonstrated a mechanistic explanation wherein inhibitory inputs could in some cases paradoxically facilitate activity [24]. The dual roles of inhibition and facilitation allow it to entrain cell activity to a signal originating in inhibitory cells.

    A relatively fine-grained, multi-formalism model is required to represent an entrainment mechanism by a simulated cell's inputs, at one scale, and the synchronization of multiple cells to plausibly generate gamma waves at a network scale. These simulations comprised local systems of ODEs, combined with a coarse PDE approximation to represent the single neuron, with event-driven techniques to connect cells into networks. To illustrate where this example fits into the spectrum of types (Fig. 2A), it is useful to focus on the way the authors modeled ion channels, as systems of ODEs. Two cross-model alternatives were used, a coarse 3-channel and a fine 11-channel representation, both ultimately derived from the underlying Hodgkin-Huxley framework. Practically, using these alternatives, helped allow for cross-model validation in the face of the greater computational complexity of the 11-channel simulations. However, from a model of explanation perspective, it is important to note that the 11-channel parameterization maps more closely to ion channel biophysics. So while both alternatives are simulations of mechanistic models, in that they are numerical solutions to systems of ODEs, the finer grained, 11-channel representation is further to the right on the Fig. 2A spectrum, toward an Analogous-mechanism Model and, ultimately, a Model Mechanism. Hence, this example exhibits different locations along the spectrum of types. It also demonstrates use of methods for moving back and forth along that spectrum.

- *Example IV-2*: More recent mechanistic explorations of gamma oscillations have focused on their possible role in the genesis of schizophrenia, where abnormalities in gamma oscillations have been demonstrated. Other clues to the biological explanation of schizophrenia have come from analogies with psychotomimetic drugs, such as ketamine. More recently, possible roles of particular molecular abnormalities have been suggested by a genome-wide association study. These many scales of causality were assessed by Neymotin et al. [25] using multiscale simulations of a mechanistic explanations to explore how alterations in one of the neural receptors at molecular scale might produce alterations in gamma oscillations in neuronal networks at the tissue scale. By using both dynamical and information theoretic measures, simulation suggested how anomalies in neuronal activity might produce disturbances in function -- disturbances in information flow. Thus the model illustrates several levels of mechanistic explanation, connecting molecular anomalies with cellular anomalies, network anomalies and information transmission disturbance.  Neurons were modeled with piecewise integrated difference equations, including inputs on the soma and dendrites, representing transmitted as well as background molecules and their receptors.  Networks of simulated neurons were composed according to a fixed relationship between three different neuron types. Simulated current injections were used to drive the network to a baseline activity, and then tuned to generate baseline theta, gamma, and theta-modulated gamma oscillations in a Local Field Potential (LFP) spanning the simulated pyramidal neurons.  The LFP oscillations provide the distinguishing phenomena. The simulated intervention mechanism consisted of turning on and off the NMDA (N-methyl-D-aspartate) inputs across 16 different cellular locations.  Because the interventions are below the network scale, instantiated by the underlying software, and mapped to the derived properties of the LFP oscillations, this model provides an excellent example of Simulation of a Mechanistic Explanation.  Further, each neuron is, itself, an example of IV, in that it is a collection of sections (soma and dendrites), each of which is a system of difference equations propagating the inputs. However, the neuronal network is designed using random connectivity, since there are no data on actual cell-to-cell connectivity.  Therefore, at this level, the model is only structurally evocative of the referent and thus approaches a Simulation of an Analogous-mechanism model (V).  By using the information theoretic measures to relate the external inputs to spike outputs, the authors were able to demonstrate an inverse relation between gamma activity and the ability of the network to transmit information, to demonstrate how gamma oscillation might underlie information processing and how gamma oscillation anomalies could underlie the abnormal information processing in schizophrenia.

*V: Simulation of an Analogous-mechanism Model*

When starting with a description of an Analogous-mechanism Model (**II**), the simulation research goal is often to translate the knowledge contained within its description into simulation output that is qualitatively and quantitatively similar to measurements of the target phenomenon.  When successful, an accurate descriptor of the work product is Simulation of an Analogous-mechanism Model.  An increasing fraction of computational explanations of phenomena reported in the literature, including some "mechanistic models" described as being "multiscale" [26], fit reasonably well under that descriptor (e.g., see [27-32]).

    Figure 4B is a snapshot of the process of building upon descriptions in **II** during workflow activities that differ from those for **IV** in important ways.  1) The scientist creates mathematical descriptions of the Analogous-mechanism Model in operation.  Continuum equations are adapted from descriptions of engineering, physical, mechanical, chemical, and/or electronic mechanisms.  An important subset of those mathematical descriptions, e.g., finite element analysis, goes beyond continuum mathematical descriptions because they also require numerical analysis techniques.  2) The mathematics is instantiated in software; features to support users are added; and solvers are selected.  Computational solutions involve solving equations subject to boundary conditions and/or initial conditions, and the implementation undergoes verification.  3) Authors undertake the iterative process of achieving qualitative and quantitative similarity between simulation output and

measurements of the target phenomenon within some tolerance. The product of that process is output from selected parameterizations of a Simulation of an Analogous-mechanism Model. The following are examples.

- *Example V-1*: Based on epidemiological studies, high-density lipoprotein (HDL) is believed to play an important role in lowering the risk of cardiovascular disease by mediating reverse cholesterol transport. Therapies that raise HDL-cholesterol, however, have been unable to confirm this hypothesis and demand a re-examination of the proposed mechanism. It is known that lipid-poor ApoA-I plays a role in initiating reverse cholesterol transport and that the drug RG7232 increases HDL-cholesterol. However, the influence of RG7232 on lipid-poor ApoA-I and reverse cholesterol transport is unclear because their direct measurement during dosing intervals is problematic. Lu et al., [28] developed an Analogous-mechanism Model and corresponding simulation to explore this response. The model is based on two other Analogous-mechanism Models, 1) a model of lipoprotein metabolism and kinetics and 2) a model of RG7232 pharmacokinetics. They are combined into a single simulation. The linked simulation goes further by additionally representing the hypothesis that the affinity of low-density lipoprotein (LDL) particles to LDL receptors is dependent on particle size or density. This hypothesis is implemented as a modified elimination rate. The resulting model describes temporal concentrations in two-compartments as coupled ordinary differential equations that are solved using the SimBiology toolbox of MathWorks. The simulation model is "analogous" in the sense that the proposed density-dependent elimination rate and compartmentalization is an analogy to chemical kinetics and chemical engineering. Parameters are estimated using a Bayesian approach that updates the parameter values from model components using the Matlab Global Optimization toolbox of MathWorks. The implementations simulate output from the linked Analogous-mechanism Model, as if it were real.

- *Example V-2*: More than 40% of astronauts who participate in long-duration missions return with ophthalmic changes similar to idiopathic intracranial hypertension. Experts posited that a microgravity-induced cephalic fluid shift elevate intracranial pressure (ICP). Feola et al. [33] hypothesized that elevated ICP would alter the peak strain environment in the optic nerve head (ONH) to cause tissue remodeling that may be contributing to the observed ophthalmic changes. They also suspected that variations in intraocular pressure (IOP) and mean arterial pressure (MAP) would affect the biomechanical strain in the OHN tissues. To explore that explanation, they implemented a finite element Analogous-mechanism Model in which a simulated structural mechanism is strongly analogous to (functions as an analog of) the ocular structure. The geometry of the analog was based on established ocular biomechanics research, and included representing coarse grain features of tissue structures known to play significant role in the observed ophthalmic changes: sclera, preliminary neural tissue, lamina cribrosa, central retinal vessel, dura mater and pia mater of the optic nerve sheath. Furthermore, an annular ring was incorporated around the scleral canal to account for the circumferential alignment of the scleral collagen fibers around the ONH. The open source package Gmsh (V2.8.3) was used to generate the 3D finite element geometry and mesh, and open source FE solver FEBio (V2.0) was used to solve for all simulations. The authors used Latin hypercube sampling of biologically plausible regions of parameter space to simulate biomechanical responses of their analog eye structure to various combinations of simulated ICPs, as well as varying IOP, MAP and simulated tissue mechanical property conditions.  Execution results showed that chronically elevated ICP coupled with interindividual differences in simulated optic nerve head mechanical properties can influence the risk for experiencing extreme optic nerve strains.  The authors inferred that individuals with both soft optic nerve or pia mater and elevated ICP would be especially at risk.

- *Example V-3*: Rosiglitazone is a PPARγ agonist, one of several approved insulin sensitizers used to treat diabetes. Despite being on the market for over a decade, the drug continues to be studied in the lab to understand the mechanism of action of this class of molecule. In Goto-Kakizaki rats, which are a rodent model of early-developing, non-obese type-2 diabetes, Gao and Jusko [31] show that rosiglitazone decreases glucose levels. To simulate how the insulin/glucose regulation might work, they built a feedback model—glucose stimulating insulin production, and insulin increasing glucose consumption. The model is analogous to other simple feedback systems, without specifying the actual, detailed, biological mechanism (e.g. intermediate steps) for glucose/insulin co-regulation.  The model also incorporates two pharmacodynamic effects of rosiglitazone that impact this feedback system: enhancing insulin sensitivity (i.e. increasing the rate of insulin-dependent loss of glucose) and inhibiting glucose production. As with many models of pharmacology, the pharmacokinetic part uses an idealized one-compartment model to fit observed drug absorption and loss. The simulation is implemented using coupled ODEs, plus analytical expressions for some of the molecules.  Given its importance to diabetes and the system under study, the component representing the time-dependent body weight of the rats was a key variable being simulated along with the molecular components.  Guided by experimental measurements (of drug, glucose, and insulin levels over time), the model was parameterized for control, low dose, and high dose rosiglitazone cases.  The match between simulation output and experiment measurements showed that the Analogous-mechanism Model explained the observations sufficiently well.  Using that model, the authors identified drug regimen design principles: specifically, to enhance insulin sensitivity in the long term (> 6 weeks), a high-dose drug is needed continuously; neither lower-dose nor shorter-term treatment succeeded in elevating the sensitivity.

- *Example V-4*: Attempts to design and build synthetic cellular memory systems using recombinases have thus far been hindered by a lack of validated computational models of a plausible mechanism representing DNA recombination.  The predictive capabilities of such models are needed to reduce the number of iterative cycles required to align experimental results with design performance requirements.  Bowyer et al. [32] developed and validated the first Simulation of an Analogous-mechanism Model for how DNA recombination might occur.  The models were constructed by extracting verified biological details from an extensive review of the experimental literature and made use of a model analogy with well-established reactions networks common to chemistry and chemical engineering.  Three essential biological details for which a consensus was lacking were included/excluded from the simulations.  The computational model consisted of a system of ODEs, each representing the concentration of a distinct biological entity, and model parameters that were optimized via the use of genetic algorithms to refine parameter values, but no details on how the model was implemented were provided.  Model predictions were compared to experimental data to determine which set of details might represent the most plausible mechanisms and thus serve as analogs of actual structural details by which DNA recombination works.  They found that including unidirectional (versus bidirectional) excision, limiting recombinase directionality factor to monomeric form in solution (versus dimer or tetramer), and integrase monomer (versus dimer) binding to DNA produced the best model match to the data. Referring to Table 1, the contextual location this Analogous-mechanism Model is implied but is not part of the implemented computational model.

**Group C: using computation to support and enhance model mechanisms**

*VI: Simulation of a Model Mechanism*

The computational mechanisms used during simulation of an Analogous-mechanism Model have nothing in common with referent mechanism's spatiotemporal entities and activities within the biological context. When a description of a Model Mechanism is available (**III**), it is feasible to change that reality by striving to actually simulate an operating, concretized software (virtual) version of the Model Mechanism. The research goal becomes twofold. 1) Create a discretized specification of the operating Model Mechanism to guide development and instantiation of a virtual mechanism. Doing so requires meeting this requirement: key portions of the virtual Model Mechanism actually operate during execution as described in **III** and contribute to the simulation of Model Mechanism features. 2) Output and/or measurements taken during simulations are qualitatively and quantitatively similar to measurements of the target phenomenon.

The workflow characterization in Figure 5A is similar to that for **IV**, except that the Model Mechanism descriptions (light blue box) are distinct in three ways. 1) Descriptions of entities and activities are discretized sufficiently to specify in software a virtual analog of the Model Mechanism that is faithful to details in **III** (e.g., see [34, 35]). 2) Evidence is presented that the entities and activities of the virtual analog are biomimetic. 3) The working hypothesis is that organized operation of software entities and activities will be capable of generating a biomimetic phenomenon.

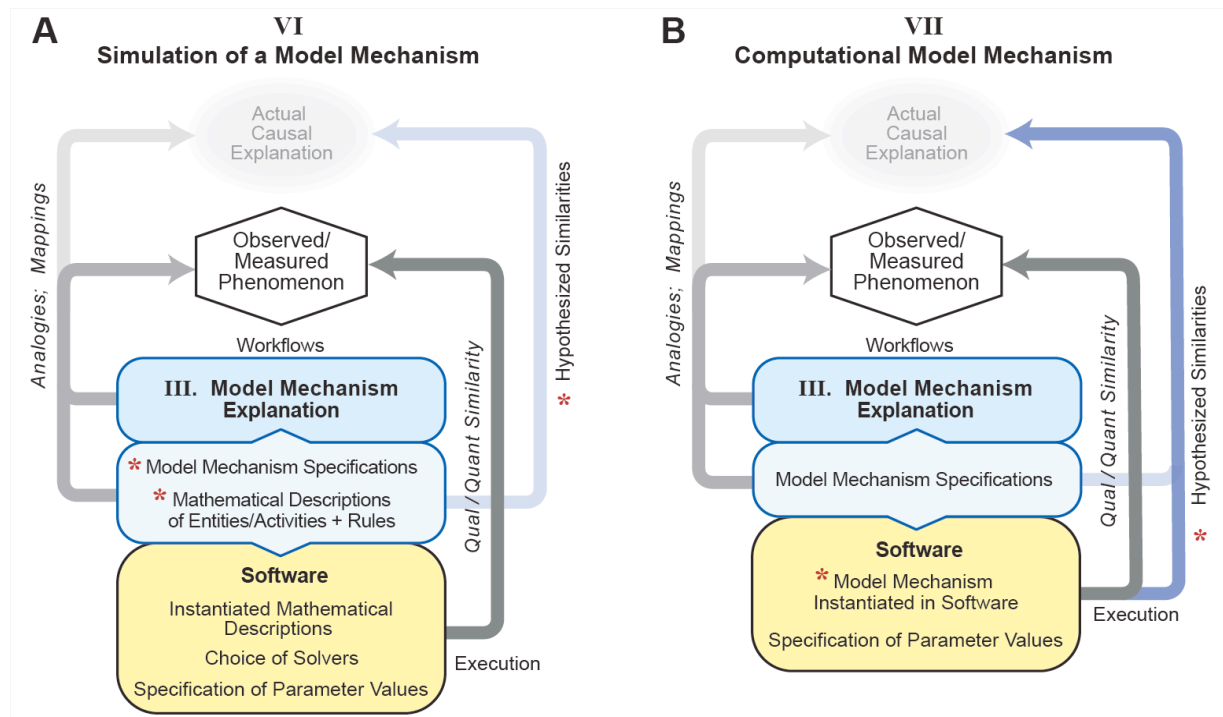

**Figure 5: Model Mechanism: from simulation to instantiation**
Illustrated are snapshots of two different work activities built upon the detailed description of a Model Mechanism in **III**. Simulation operation is not illustrated. A requirement for both is that output matches target phenomenon measurements within some tolerance. (A) Red asterisks identify characteristics that distinguish a Simulation of a Model Mechanism from **V**. Agent-oriented simulation methods are often utilized. To the extent feasible, envisioned entity activities are described using probabilistic and/or deterministic rules. Often, however, to simplify technical implementation challenges, behaviors of all or some Model Mechanism activities during execution are simply described using continuous mathematics, as in **V**, using physically grounded parameterizations. So doing prevents some or all of the software mechanisms during execution from meeting the definition of mechanism. (B) The red asterisk identifies a characteristic that distinguishes this implementation from **VI**. Authors strive to use the Model Mechanism specifications to instantiate a software analog of the entire Model Mechanism. The product is a Virtual Model Mechanism. To build credibility, authors demonstrate that a parameterized variant of the Virtual Model Mechanism has met the five requirements listed in the text. A distinguishing element is that features of the software mechanism during execution are observable, measurable, and hypothesized to have analogous biological counterparts (blue arrow).

To achieve computational efficiencies and/or fine grain details, such as receptor trafficking, signaling networks, and molecular diffusion, influences of some entities and activities within the larger Model Mechanism are often described using a combination of rules and continuous mathematics, as in **V**, rather than being implemented as discrete biomimetic entities and activities. Doing so causes the software mechanisms during execution to fall short of the definition of mechanism [36]. Nevertheless, an accurate descriptor of the work product is a Simulation of a Model Mechanism. The following are examples.

- *Example VI-1*: Simulations of Model Mechanisms are being used to help design and improve therapeutic interventions in disease [37-39]. For example, they are providing improved insight into possible failure modes of current treatments strategies for Tuberculosis (TB). Building on their multilevel, multi-attribute Model Mechanism of an immune response to TB, Linderman et al. [35] explored simulations of consequences of potential new pharmacological interventions on six different model entities and activities, including simulating immunomodulation by a cytokine; the consequences of oral and inhaled antibiotics; and the effect of vaccination. Inline with the features of a biological mechanism (Table 1), their Model Mechanism identifies a phenomenon, the immune response of TB as indicated by granuloma formation and function. Components are represented at different spatial and temporal scales describe, starting with an agent-based analogy of cell behavior (macrophages and T cells) across a cross-section of lung tissue. Through rule-based probabilistic interactions, cell behavior is simulated in response to a bacterial environment. At the lowest levels of simulation hierarchy, ordinary differential equations were solved within each cell agent to simulate receptor/ligand binding, trafficking, and intracellular signaling. Partial differential equations were solved to simulate consequences of molecular diffusion. By linking their Simulation of a Model Mechanism for TB to ordinary differential equation-based pharmacokinetic and pharmacodynamic models, the authors simulated plausible consequences of the Model Mechanism's behavior during exposure to antibiotics. While simulations rely on some model compartments that are Analogous-mechanism Models, the whole system is arguably a Model Mechanism. It is biomimetic and represents an interconnected biological mechanism of granuloma formation and immune response that extends from molecular to organ levels.

- *Example VI-2*: A decade ago, several laboratories sought improved models of explanation for vascular patterning defects observed in diabetic retinopathy and tumor angiogenesis. Evidence suggested that an explanatory mechanism would involve disruption of 1) notch-driven specialization of endothelial cells into leading tip cells and following stalk cells, and 2) a feedback loop that links VEGF-A tip cell induction with delta-like 4 (Dll4)-notch-mediated lateral inhibition. Bentley et al. [40] constructed a hierarchical Simulation of a Model Mechanism to explore the phenomenon of angiogenesis by connecting Analogous-mechanism Models of these processes into a large biomimetic system. The components included endothelial cell agents and membrane agents with multiple cell agents arranged as a cylindrical capillary with each cell having membrane agents distributed at the periphery. The study explored how different simulated VEGF environments and filopodia dynamics would affect simulations of notch-mediated selection of tip cells. A staged simulation (temporally and spatially) first relied on a rule-based evaluation of membrane processes for filopodium retraction or extension or notch response to VEGF. In following, the spatial sum of protein levels was calculated and redistributed within the endothelial cells and membrane agents. The modeling paradigm closely follows that of a Model Mechanism, where features reflect those of a biological mechanism (Table 1). An important observation of the simulations was that, by removing information that could influence simulated cell biasing, the simulated Dll4-notch lateral inhibition mechanism could generate an alternating pattern of cell fates characteristic of normal tip cell selection. The authors inferred from simulation results that abnormal patterning could be attributed to the dynamics of this particular sub-system, rather than any uncontrolled bias.

*VII: A Virtual Model Mechanism*

This characterization differs from that in **VI** in one important way. All features of a Model Mechanism instantiated in software meet the definition of mechanism during operation and may include all of the features in Table 1. To do so, five requirements are specified early in the workflow to guide software engineering, mechanism instantiation, and simulation refinements. 1) Evidence is presented that entities and activities of the virtual mechanism are biomimetic. 2) Features of the software Model Mechanism during execution are measurable. 3) Biomimetic phenomena are generated during execution. 4) Measurement of features of one or more simulation solutions match or mimic measurements of the target phenomenon within some tolerance (e.g., see [34, 41]). 5) Arguments can be presented that, during execution, the Model Mechanism will have a biological counterpart (blue arrow in Figure 5B). Here are three examples.

- *Example VII-1*: Enhanced mechanism-based explanations are needed to anticipate, prevent, and reverse the liver injury caused by acetaminophen and other drugs. A characteristic acetaminophen phenomenon—the target phenomenon for this example—is that hepatic necrosis begins adjacent to central veins in hepatic lobules and progresses upstream. The prevailing (mechanism-oriented spatiotemporal) explanation (PE) is that location dependent differences in reactive metabolite formation within hepatic lobules (called zonation) are necessary and sufficient requisites to account for the phenomenon.  Progress has been stymied because challenging that hypothesis in mice would require sequential intracellular measurements at different lobular locations within the same mouse, which is infeasible.  Smith et al. [34] circumvent that impediment by performing experiments on virtual Mouse Analogs, where each is equipped with an in silico liver that achieved multiple validation targets.  Components and spaces at all levels of granularity are written in Java, utilizing the MASON multi-agent simulation toolkit.  An accurate causal model of the PE that exhibits all Table 1 features was instantiated and parameterized so that, upon dosing with objects representing acetaminophen, metabolism and pharmacokinetic validation targets were achieved.  However, the authors demonstrated that the PE failed to achieve the target phenomenon.  Two parsimoniously more complex variants also failed to achieve the target phenomenon, but a fourth variant met stringent tests of sufficiency.  Execution of that forth Computational Model Mechanism provided a multilevel biomimetic causal explanation of key temporal features of acetaminophen hepatotoxicity in mice including the target phenomenon.  The authors argue that the causal explanation provided during execution is strongly analogous to the actual causal mechanism in mice.

- *Example VII- 2*: Inflammation is not the result of one cell or molecule acting alone.  It is a multicellular process that can be highly localized and yet also have diffuse actions. One of the keys to understanding tissue-level morphogenesis and spatially localized or heterogeneous processes such as inflammation is to explicitly study the spatial component - how the cells are arranged in the tissue and the influences that they have on each other. Thus, to gain insight into the pathogenesis of gastrointestinal inflammatory diseases, Cockrell et al. [42] developed a multi-level, discrete-event Model Mechanism that is used to study scenarios of how simulated cellular and molecular pathways may govern morphogenesis and inflammation in healthy and disease ileal mucosal dynamics. The system includes individual agents representing five different cell types, each with multiple independently acting instantiations at different physical locations. Cell agents have specific behaviors (proliferation, death, anoikis, etc.) and can influence each other's decision-making process.  Inside each agent, there is also a simulated signaling network. The system uses algebraic rules to simulate most of the different components, including a representation of extracellular paracrine signaling between cells (with the addition of a grid-based partial differential equation to simulate consequences of diffusion), the dynamics of the simulated intracellular signaling networks, and (using the current values of key intracellular signaling components as a basis) the likelihood of cell agents exhibiting each possible behavior. By simulating cell behavior in a virtual world that is analogous to biological microenvironments, the system can generate measurable phenomena (predictions) at multiple levels.  Simulations provide insight into plausible pathological processes, including crosstalk between morphogenesis and inflammation, and the effects of cell death on tissue health.

- *Example VII-3*: Changes to savanna ecosystems related to climate change and land use practices are linked to fluctuations in savanna bird community structures, functional traits, and risk of extinction.  Better, more insightful models of explanation are needed to support policy changes.  However, detailed species-specific data for a given ecosystem are often limited. As a method test case for overcoming such limitations, Scherer et al. [43] used an agent-oriented approach (implemented in NetLogo) that merged trait-based and individual-based simulation methods to predict how different bird functional types might change in response to concurrent alterations to savanna rangeland from a combination of climate change and land use.  The entire simulated ecosystem operates during execution as a Model Mechanism.  Contained within are all of the features listed in Table 1.  The system includes a spatial and stochastically varying set of entities representative of the type of individual, home range, vegetation, landscape, and environment.  Each entity was characterized by a set of state variables, examples of which include age and reproductive status, or grasses, shrubs, or trees.  Executions advance in uniform steps that map to an interval of up to 100 years, and progress by randomly selecting, calculating, and updating properties that control the spatial composition and configuration of simulated habitat and animals.  Simulation results provided possible explanations for why simulated extinction risks for simulated larger- bodied insectivores, omnivores, and small-bodied species were impacted differently by changes in simulated shrub-grass ratio and clumping intensity of shrub patches.  Such predictions could prove essential for identifying better policies for conservation management.

## RELEVANT INFORMATION, MULTIPLE SOURCES

Essential relevant information from a variety of sources is needed to establish and enhance the credibility of improved insights that may be derived from **IV–VII**.  The Figure 2B and 2C spectra characterize two important sources.  The three Figure 6 spectra identify additional information sources and types.  The Figure 6 spectra are more closely linked to methodology than are the workflow characterizations in Figures 3-5.  Having essential information available enables authors and readers to identify approximate locations on all five spectra, which improves clarity and brings into focus the characteristics that distinguish among **IV–VII**.

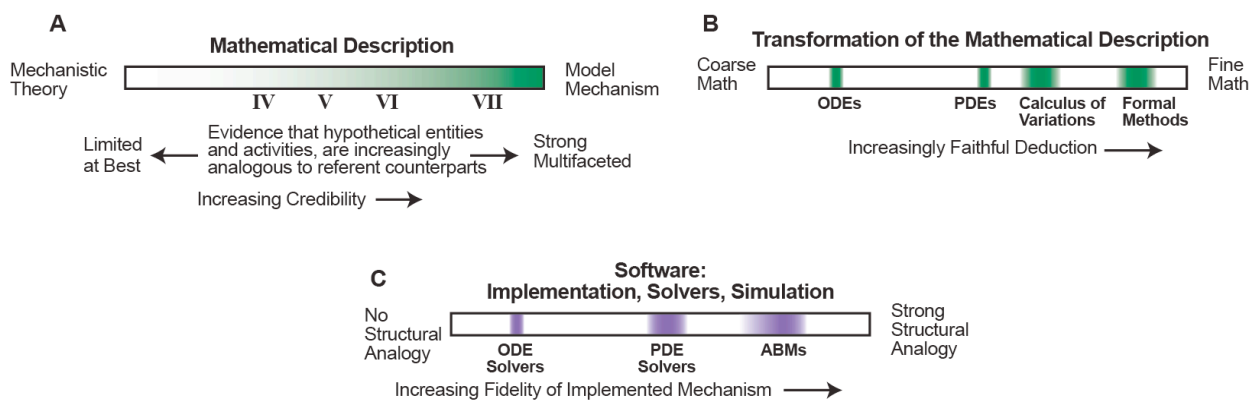

**Figure 6: Characteristics three sources and types of relevant information**
These three spectra are distinct from those in Figure 2.  They bring into focus characteristics of methods and approach used in workflows that distinguish among **IV–VII**.  (A) The relationship between **I**, **II**, or **III** and the corresponding mathematical description must be clear.  (B) Expanding a model or combining it with other models [44, 45] (e.g., to create a hybrid model) is a strategy used to improve explanatory descriptions.  The choice of mathematical description used for the expansion or combination can influence faithfulness of deductive transformations.  Four examples of commonly used mathematical model types are shown to illustrate that different types occupy different relative locations.  Some mathematical model types cannot be easily modified and remain faithful to the target phenomenon while also preserving the original meaning(s) of the model's terms and model-to-target mappings provided in the explanatory descriptions.  (C) This spectrum illustrates that implementation decisions (primarily within the yellow boxes in Figures 4 and 5) influence the fidelity of the biomimesis that can be built into the simulations during execution. Stronger analogies between the biology and model mechanisms during execution are expected to improve clarity, credibility and scientific usefulness.

Spectrum 6A characterizes the mathematical descriptions used in **IV–VI**.  Information is lost during derivation from the primarily prosaic description (including induction from data) in **II** and **III** to mathematical descriptions.  Clarity about what is and is not lost can influence credibility.  For example, the assumption behind Simulation of an Analogous-mechanism Model is that, if the model were made real, then some version of the phenomenon generated during operation would mimic the referent phenomenon.  In most reports, the focus is primarily on mimicking the referent phenomenon, and much less so on the model's entities, activities, and their organization during phenomenon generation.  Consequently, it is often the case that mathematical descriptions are imbalanced, which can limit clarity and credibility.

Spectrum 6B is about (primarily deductive) transformations of the descriptions in **I–III**.  The research goal of improving mechanism-oriented explanations often involves inferring plausible biological details from explorations of the model's behavior and then seeking transformations (ways to change computational features) that provide improvement.  Formal Methods refer to the computer science (and mathematics) that allows such transformations to be rigorous enough to reason over, i.e., to make them purely deductive.  Particular types of mathematical models (e.g., ODEs) cannot be easily modified without breaking the extent to which the model represents the description in **II** or **III** and maps to the target phenomenon.  Faithful deduction over a simulation, including modifications that are faithful to the target phenomenon, are those that preserve the original meaning(s) of the model's terms and model-to-target phenomenon mappings (for example [44]).  The expectation is that credibility of **IV–VII** will increase as faithfulness to deductive transformations from mathematical descriptions increases.

Spectrum 6C illustrates that implementation decisions influence the fidelity of biomimesis built into a simulation during execution.  We anticipate that the deeper the insight, the stronger the analogy between the biology's mechanisms and simulation's mechanisms.  Thus, credibility will increase by increasing structural analogies between implementations simulating the target phenomenon and the biological system generating the target phenomenon.

Moving rightward on spectra 2B and 2C involves incorporating deeper (validated) insight into an expanding variety of interconnected biological processes and phenomena.  Mechanism-oriented models that are developing that insight into an expanding variety of phenomena will be moving rightward on the Figure 6 spectra.  As a consequence, implementations must change during each move to the right.  During those changes, information that can influence—bias—simulation output can be lost and/or added.  Documenting those influences enhances credibility.  The absence of such documentation risks creating a barrier to credibility, thus limiting scientific usefulness.

**WORKFLOW, PROVENANCE, AND HYBRID MODELS**

Most biological scientists and clinicians have a general appreciation for, and understanding of, the workflow, the systems utilized, and methods employed in wet-lab research.  When they read a research article reporting results of experiments, that knowledge influences their assessment of credibility.  Biological scientists and clinicians outside of the simulation field may be drawn to (and may consider reading) a simulation-focused research report due to the prospect for improved explanatory insight or practical utility.  However, they do not have a corresponding appreciation for, or understanding of, the workflow, the systems utilized, or the methods employed.  Thus, there is a significant risk that missing information and lack of clarity will erode the reader's assessment of the credibility of arguments presented, and of simulation approaches in general.

The credibility of inferences about a phenomenon based on results of wet-lab experiments depends on having easy access to the experiment's provenance [46], i.e., full context of the experiment along with adequate descriptions of methods, materials, and other important workflow details.  Removing or distancing observations and/or data from the experiment's provenance abstracts away both information and knowledge, thus weakening justifications for their application or use elsewhere. By analogy, the credibility of explanations provided by simulations for how a phenomenon may be generated depends on use context, and includes having easy access to the provenance of **IV–VII** [47].  Provenance begins with **I–III** and includes the full context of the simulation activities.  Also by analogy, unlinking an element (e.g., a mathematical description or software implementation detail) from the information and knowledge provided by the original use context and provenance for application or reuse elsewhere can weaken or eliminate justifications for the intended application or reuse, thus eroding credibility and limiting scientific usefulness.

It is now common to encounter biology simulation research reports that seek merged explanations of two or more phenomena or a description of phenomena across multiple biological levels or scales.  The software instantiations, commonly referred to as hybrid models, require means for the different, originally separate and independent mechanism-oriented models to interact during execution.  Those means include adding software features and making changes to the previously independent implementations.  Describing the product of that process as a hybrid alerts readers to expect the merged system to behave in new ways.  Some behaviors will be intended, but others may be unintended.  The situation is somewhat analogous to combining two reagents during a wet-lab protocol when, under some conditions, doing so risks an adverse interaction.  The importance of providing clear details is obvious.

**CONCLUDING REMARKS**

Although credibility and clarity are often correlated, other factors can have even greater influence on explanatory credibility.  Each element in the **I–VII** characterizations will "resonate" differently with different scientists, clinicians, and stakeholders.  Here are three examples: 1) The evidence selected to support a description of an Analogous-mechanism Model (**II**) may resonate well with engineers and system biologists but less so with oncologists.  2) For a particular characterization, the interpretations offered by authors in the context of selected simulation results will likely resonate differently with scientists approaching the problem from basic science and clinical perspectives.  3)

The extent to which a particular set of mathematical expressions or software engineering methods resonates with a simulation researcher will likely have a significant impact on that person's determination of whether a particular computational mechanism-oriented model is sufficiently mechanistic or not, which, in turn, may impact that person's assessment of credibility. There are, of course, other influences and even larger issues to consider. For example, the interpretation of what is happening within all the above workflows is part of the philosophy of science. We put these important influences aside for now; they are beyond the scope of this overview.

Increasing complexity in pursuit of mechanism-oriented models that improve explanatory credibility is an explicit strategy within biology simulation research (e.g., see [27, 45, 47]). For the larger community of biologists, a priority is achieving deeper, more useful explanations of phenomena that facilitate advancing both science and health. The scientific usefulness of biology simulation as a discipline will become more evident to the larger community as credible multi-phenomena explanations become available. Achieving credible multi-phenomena explanations requires moving rightward on all spectra in Figures 2 and 6, but doing so requires increasing support from the larger biology community. Improving clarity, semantics and otherwise, is a necessary and essential small step to achieving that increased support.

By characterizing **I–III** and **IV–VII** we demonstrate how semantic clarity can be improved even as the complexity of those models of explanation increases. These categories of types of models and simulations may serve as a foundation for a clear ontology of mechanism-based simulation research in biology.

## ACKNOWLEDGMENTS

We thank Glen E.P. Ropella for providing content suggestions; Andrew Smith and Ryan Kennedy for constructive criticism during manuscript development; and Mitzi Baker editorial input.  This work was supported in part by the National Institutes of Health: R01GM104139 (AE), R01HL101200 (FMG), R01EB022903 (WWL); The National Science Foundation: NSF:CAREER 1452728 (EAS); InSilico Labs LLC (LM); and the UCSF BioSystems group (CAH).  Authors' contributions: C.A.H. managed manuscript preparation, content organization, development, and editing; A.E., W.W.L., F.M.G, E.A.S., M.K.T., L.M. and C.A.H. contributed to the development of the presented ideas; L.M. collected the data in Figure 1; C.A.H. created Figures 2-6.  The authors declare no competing interests.

Preprint: Jan. 05, 2018                                                                                        29

## Further Reading